\documentstyle[12pt]{article}
\newlength{\mathspace}
\def\np#1{ Nucl. Phys. B#1}
\def\pr#1    { Phys. Rev. D#1 }
\def\pl#1{ Phys. Lett. B#1}

\def\ijmp#1  { Int. Jour. Mod. Phys. A#1 }
\def\mpl#1   { Mod. Phys. Lett. A#1 }
\def\begineq{\begin{equation}}
\def\endeq{\end{equation}}
\def\eqabegin{\begin{eqnarray}}
\def\eqaend{\end{eqnarray}}
\def\nn{\nonumber}

\begin{document}
\baselineskip=0.7cm
\setlength{\mathspace}{2.5mm}



\begin{titlepage}

    \begin{normalsize}
     \begin{flushright}
  		 CTP-TAMU-06/98\\
                 SINP-TNP/98-05\\
                 hep-th/9802080\\
     \end{flushright}
    \end{normalsize}
    \begin{LARGE}
       \vspace{1cm}
       \begin{center}
         {An SL(2, Z) Multiplet of Type IIB Super Five-Branes}\\ 
       \end{center}
    \end{LARGE}

  \vspace{5mm}

\begin{center}
           
             \vspace{.5cm}

            J. X. L{\sc u}\footnote[2]{Research supported by NSF Grant 
  	    PHY-9722090, e-mail address: 
            jxlu@phys.physics.tamu.edu}

            \vspace{2mm}

            {\it Center for Theoretical Physics}\\
            {\it Physics Department, Texas A \& M University}\\
            {\it College Station, TX 77843, USA}

            \vspace{2mm}
            
            and

            \vspace{2mm}
 
            Shibaji R{\sc oy}
           \footnote{E-mail address: roy@tnp.saha.ernet.in} 

                 \vspace{2mm}

{\it Saha Institute of Nuclear Physics}\\
        {\it 1/AF Bidhannagar, Calcutta 700 064, India}\\

      \vspace{2.5cm}

    \begin{large} ABSTRACT \end{large}
        \par
\end{center}
 \begin{normalsize}
\ \ \ \
It is well-known that the low energy string theory admits a non-singular
solitonic super five-brane solution which is the magnetic dual to the
fundamental string solution. By using the symmetry of the type IIB
string theory, we construct an SL(2, Z) multiplet of magnetically charged
super five-branes starting from this solitonic solution. These solutions
are characterized by two integral three-form charges $(q_1,\,q_2)$ and
are stable when the integers are coprime. We obtain an expression for
the tension of these $(q_1,\,q_2)$ five-branes as envisaged by Witten.
The SL(2, Z) multiplets of black strings and black fivebranes and the 
existence of similar magnetic dual solutions of strings in type II string theory
in $D < 10$ have also been discussed.
\end{normalsize}

\end{titlepage}
\vfil\eject

String theories in the long wavelength limit are described by various kinds 
of supergravity theories in $D = 10$. The supergravity equations of motion
are well-known [1--3] to admit various black $p$-brane solutions which
are essentially the black hole solutions of the dimensionally reduced
low energy string effective action spatially extended to the ten
dimensional theory. These solutions are usually characterized by two 
parameters related to the charge and the mass of the black $p$-branes.
In the extremal limit as usual in the Reissner-Nordstrom black hole,
the charge and the ADM mass of the $p$-branes are related to each other
and the solutions become supersymmetric saturating the BPS bound. The
extremal solutions of string effective actions are particularly
interesting since their masses and the charges can be calculated exactly
due to certain non-renormalization theorems of the underlying 
supersymmetric theories. Thus although these solutions are obtained from
the low energy effective theory, they are quite useful to identify
certain non-perturbative symmetries of string theory.

Low energy effective action of any string theory admits a fundamental
string solution [4] and its magnetic dual the non-singular solitonic
five-brane solution [5--9]. These are the extremal limit, as we have
mentioned, of the black string and black five-brane solutions of the
corresponding supergravity equations of motion. Since it is known that
the equations of motion of type IIB supergravity theory is invariant
[10] under an SL(2, R) group one can construct a more general string
like as well as five-brane solutions using this symmetry group.
We should like to mention that the SL(2, R) symmetry of type IIB
string theory is a non-perturbative symmetry. It transforms the
string coupling constant in a non-trivial way and therefore mixes up
the perturbative and non-perturbative effects of Type IIB string theory. 
A discrete subgroup of this SL(2, R) group has been conjectured to be the exact
symmetry group of the quantum type IIB string theory [11]. Using this
symmetry, Schwarz [12] has constructed an SL(2, Z) multiplet of string
like solutions in type IIB string theory starting from the fundamental
string solution. Both the string tension and the charge were shown
to be given by the SL(2, Z) covariant expressions. Since the tension 
and the charge of these extremal solutions remain unrenormalized, it
provides a strong support in favor of the conjecture that SL(2, Z)
is an exact symmetry group of the quantum theory.

Given the symmetry of the type IIB theory, it is natural to expect
that the solitonic five-brane solution should also form an SL(2, Z)
multiplet as pointed out by Schwarz in ref.[12]. In this paper,
we construct\footnote[1]{Earlier attempt for the construction
of SL(2, Z) multiplet of five-brane solution of type IIB string theory 
were made in [13,14], but the solutions described there are incomplete.} 
an infinite family of magnetically charged super 
five-branes, permuted by SL(2, Z) group, starting from the solitonic
five-brane solution of string theory. These solutions are characterized
by a pair of integers corresponding to the magnetic charges associated
with the two three-form field strengths present in the NSNS and RR 
sector of the spectrum. When these two integers $(q_1,\,q_2)$ are
relatively prime to each other, the five-brane solutions are shown
to be stable and can be regarded as bound state [15] configuration
of $q_1$ solitonic five-branes with $q_2$ D5-branes [16]. The magnetic
charge as well as the five-brane tension are shown to be given by
SL(2, Z) covariant expressions. This provides more evidence that
SL(2, Z) is indeed an exact symmetry group of the quantum theory.
The expression for the tension of these $(q_1,\,q_2)$ super five-branes
has been envisaged by Witten [15] some time ago. We then discuss
that similar magnetic dual solutions of strings also exist in lower dimensional
type II theory.

The low energy effective action common to any ten dimensional string
theory has the form:
\begineq
S = \int\,d^{10} x \sqrt {-g} 
\left[R - \frac{1}{2} \nabla_\mu \phi \nabla^\mu \phi - \frac{1}{12}
e^{-\phi} H_{\mu\nu
\lambda}^{(1)} H^{(1)\,\mu\nu\lambda}\right]
\endeq
Here $g = ({\rm det}\,g_{\mu\nu})$, $g_{\mu\nu}$ being the canonical
metric which is related to string metric by $G_{\mu\nu} = e^{\phi/2 }
g_{\mu\nu}$. $R$ is the scalar curvature with respect to the canonical
metric, $\phi$ is the dilaton and $H_{\mu\nu\lambda}^{(1)}$ is the field
strength associated with the Kalb-Ramond antisymmetric tensor field
$B_{\mu\nu}^{(1)}$. These are the massless modes which couple to any 
string theory. For type II strings these massless modes belong to the
NSNS sector of the spectrum. The equations of motion following from
(1) are:
\eqabegin
& & \nabla_\mu\left(e^{-\phi} H^{(1)\,\mu\nu\lambda}\right)\,\,\,
=\,\,\,0\\
& &\nabla^2 \phi + \frac{1}{12} e^{-\phi} \left(H^{(1)}\right)^2\,\,\,
=\,\,\,0\\
& & R_{\mu\nu} - \frac{1}{2} \nabla_\mu \phi \nabla_\nu \phi -
\frac{1}{4} e^{-\phi} H_{\mu\rho\sigma}^{(1)} H_\nu^{(1)\,\,
\rho\sigma} + \frac{1}{48} e^{-\phi} \left(H^{(1)}\right)^2 
g_{\mu\nu}\,\,\,=\,\,\,0
\eqaend
These low energy field equations can be solved by using certain ansatz
on the metric and the field strength $H_{\mu\nu\lambda}^{(1)}$. By
demanding that the metric be static, spherically symmetric which 
becomes flat asymptotically with a regular horizon, one can obtain
both the electrically charged black string solution and the magnetically
charged black five-brane solution from (2) -- (4) as given below [3]
\eqabegin
ds^2 &=& -\left(1 - \frac{r_+^6}{r^6}\right)\left(1 - \frac{r_-^6}
{r^6}\right)^{-1/4} dt^2 + \left(1 - \frac{r_-^6}{r^6}\right)^{3/4}
(dx^1)^2\nn\\
& &\qquad\qquad+\left(1 - \frac{r_+^6}{r^6}\right)^{-1} \left(1 - \frac{
r_-^6}{r^6}\right)^{-11/12} dr^2 + r^2 \left(1 - \frac{r_-^6}{r^6}\right)^
{1/12} d\Omega_7^2\nn\\
e^{2\phi} &=& \left(1 - \frac{r_-^6}{r^6}\right),\qquad  H^{(1)}\,\,\,
=\,\,\,6\left(r_+ r_-\right)^3 \ast e^{\phi} \epsilon_7
\eqaend
for the string solution and
\eqabegin
ds^2 &=& -\left(1 - \frac{r_+^2}{r^2}\right)\left(1 - \frac{r_-^2}
{r^2}\right)^{-3/4} dt^2 + \left(1 - \frac{r_-^2}{r^2}\right)^{1/4}
\delta_{ij} dx^i dx^j\nn\\
& &\qquad\qquad+\left(1 - \frac{r_+^2}{r^2}\right)^{-1} \left(1 - \frac{
r_-^2}{r^2}\right)^{-3/4} dr^2 + r^2 \left(1 - \frac{r_-^2}{r^2}\right)^
{1/4} d\Omega_3^2\nn\\
e^{-2\phi} &=& \left(1 - \frac{r_-^2}{r^2}\right),\qquad  H^{(1)}\,\,\,
=\,\,\,2\left(r_+ r_-\right) \epsilon_3
\eqaend
for the five-brane solution. Here $i,\,j = 1, 2, \ldots, 5$. 
$d\Omega_7^2$ and $d\Omega_3^2$ above are the metric on the unit 
seven dimensional and three dimensional spheres respectively. 
$\epsilon_7$ and $\epsilon_3$ are the corresponding volume forms.
The `$\ast$' denotes the Hodge duality transformation. $r_+$ and $r_-$ 
are the two
parameters related to the mass and the charge of the solutions.
Thus eqs. (5) and (6) represent the two parameter family of black
string and black five-brane solutions of string theory with an event
horizon at $r = r_+$ and an inner horizon $r = r_-$ (where $r_+ \geq
r_-$). It is clear from the form of $H^{(1)}$ in (5) that the string
solution is electrically charged whereas from (6) we note that the
five-brane solution is magnetically charged. We would like to 
mention that if the supergravity action contains a general 
$(d + 1)$-form field strength then the magnetically charged 
solution can be obtained from electrically charged solution by using 
the duality
transformation
$\phi \rightarrow -\phi$,
$e^{-\phi} \ast H^{(1)} \rightarrow H^{(1)}$ and $d \rightarrow 8-d$.
Thus the five-brane solution can be obtained from the string solution
using this duality transformation.
Since, 
$e^\phi$ is the string coupling constant, the magnetically charged 
five-brane solution is a non-perturbative solution of weakly coupled
string theory. Note that the solutions (5) and (6) are written
assuming that the dilaton vanishes in the asymptotic limit, but we
will restore the asymptotically constant value of the dilaton when
we write the more general type IIB solution. Finally, we note that
for $r_+ > r_-$, the solutions are non-extremal and therefore non-BPS,
but when $r_+ = r_-$ the solutions become supersymmetric saturating
the BPS limit.

In the extremal limit the BPS saturated string solution given in (5)
(with $r_+ = r_-$), was constructed previously by Dabholkar et. al.
[4]. By going to the isotropic coordinate $\rho^6 = r^6 - r_-^6$, we
can rewrite the metric (5) in the extremal limit as,
\begineq
ds^2 = \left(1 + \frac{r_-^6}{\rho^6}\right)^{-3/4}\left[-(dt)^2 +
(dx^1)^2\right]
+\left(1 + \frac{r_-^6}{\rho^6}\right)^{1/4}\left(d\rho^2 + \rho^2 d
\Omega_7^2\right)
\endeq 
with $e^{-2\phi} = \left(1 + \frac{r_-^6}{\rho^6}\right)$. This is
precisely the solution discussed in ref.[4] and clarifies the relation
with the solution described in [3]. It is
clear from (7) that in terms of the string metric $G_{\mu\nu} =
\left(1 + \frac{r_-^6}{\rho^6}\right)^{-1/4} g_{\mu\nu}$, the above
metric becomes flat transverse to the string direction. Also note that the
solution (7) is singular since for $\rho \rightarrow 0$, the radius
of $S^7$ vanishes and the curvature blows up as $\rho^{-2}$. This is the
reason that string like solution has been constructed [4] by coupling
the supergravity action to a macroscopic string source. This BPS
saturated singular string like solution is also known as the
fundamental string solution. By using the symmetry of the full type
IIB string theory including the RR sector, Schwarz has constructed an 
infinite family of string like solutions starting from this fundamental
string solution. In ref.[17], we have pointed out that a similar infinite
family of string solutions also exist in $D < 10$ type II string theory.

Let us next look at the black five-brane  solution of string theory given
in (6). Note that in general the solution is invariant under the symmetry
group R $\times$ E(5) $\times$ SO(4) where E(5) is the five dimensional 
Euclidean group. At the extremal limit, the solution acquires an extra
boost symmetry and thus the symmetry group becomes P(6) $\times$ SO(4),
where P(6) is the six dimensional Poincare group. So, only at the 
extremal limit the solution describes the BPS super five-brane and takes
the following form:
\eqabegin
ds^2 &=& \left(1 - \frac{r_-^2}{r^2}\right)^{1/4}\left[-(dt)^2 +
\delta_{ij} dx^i dx^j\right]
+\left(1 - \frac{r_-^2}{r^2}\right)^{-7/4} dr^2 + r^2 \left(1 - 
\frac{r_-^2}{r^2}\right)^{1/4} d\Omega_3^2\nn\\
Q &=&2r_-^2,\qquad e^{-2\phi}\,\,\, =\,\,\, \left(1 - 
\frac{Q}{2 r^2}
\right),\qquad H^{(1)}\,\,\,=\,\,\,Q \epsilon_3
\eqaend
In terms of the isotropic coordinate $\rho^2 = r^2 - r_-^2$ the solution
can be written as,
\eqabegin
ds^2 &=& \left(1 + \frac{r_-^2}{\rho^2}\right)^{-1/4}\left[-(dt)^2 +
\delta_{ij} dx^i dx^j\right]
+\left(1 + \frac{r_-^2}{\rho^2}\right)^{3/4}\left(d\rho^2 + \rho^2 
d\Omega^2_3\right)\nn\\ 
Q &=& 2r_-^2,\qquad e^{2\phi}\,\,\, =\,\,\, \left(1 + 
\frac{Q}{2 \rho^2}
\right),\qquad H^{(1)}\,\,\,=\,\,\,Q \epsilon_3
\eqaend
In the string frame $G_{\mu\nu} = \left(1 + \frac{r_-^2}{\rho^2}\right)^
{1/4} g_{\mu\nu}$, the metric in (9) reduces to
\begineq
ds^2 = \left[-(dt)^2 +
\delta_{ij} dx^i dx^j\right]
+\left(1 + \frac{r_-^2}{\rho^2}\right)\left(d\rho^2 + \rho^2 
d\Omega^2_3\right)
\endeq
Note that unlike the string solution, the super five-brane solution is
regular as $\rho \rightarrow 0$ since in this case the radius of $S^3$
is finite ($r_-$) and the curvature also remains finite ($r_-^{-2}$)
as $\rho \rightarrow 0$.

We would like to point out that the low energy string effective action
we have considered in (1) can be regarded as a special case of more
general type IIB action when the RR fields are included. Let us recall
that the massless states of type IIB string theory in the bosonic sector
consist of a graviton ($G_{\mu\nu}$), a dilaton ($\phi$) and an 
antisymmetric tensor field ($B_{\mu\nu}^{(1)}$) as NSNS gauge fields 
whereas in the RR sector it has another scalar ($\chi$), another
antisymmetric tensor field ($B_{\mu\nu}^{(2)}$) and a four-form gauge
field ($A_{\mu\nu\rho\sigma}^+$) whose field strength is self-dual.
It is well-known that the equations of motion of type IIB supergravity
theory is invariant under an SL(2, R) group known as the supergravity
duality group [10]. The four-form gauge field is a singlet under this
duality group and it couples to a self-dual three-brane whose form
has been derived in ref.[3,18]. Since we are interested in the five-brane
solution we set the corresponding five-form field strength to zero
in what follows. In this case the type IIB supergravity equations of 
motion can be derived from the following covariant action [19],
\eqabegin
\tilde {S}_{{\rm IIB}} &=&
\int\,d^{10} x \sqrt {- G} \left[ e^{-2 \phi}
\left(R + 4 \nabla_{\mu} \phi \nabla^\mu 
\phi
- \frac{1}{12} H_{\mu\nu\lambda}^{(1)} H^{(1)\, \mu
\nu\lambda}\right)\right.\nn\\
& &\qquad \left. - \frac{1}{2} \nabla_\mu \chi \nabla^
\mu
\chi -\frac{1}{12}\left(H_{\mu\nu\lambda}^{(2)} 
+ \chi H_{\mu\nu\lambda}^{(1)}\right)\left(
H^{(2)\, \mu\nu\lambda} + \chi H^{(1)\,
\mu\nu\lambda}\right)\right]
\eqaend
We can rewrite the action (11) in the Einstein frame as follows:
\eqabegin
 {S}_{{\rm IIB}} &=&
\int\,d^{10} x \sqrt {- g} \left[ 
R - \frac{1}{2} \nabla_{\mu} \phi \nabla^\mu 
\phi - \frac{1}{2} e^{2\phi} \nabla_\mu \chi \nabla^\mu \chi\right.\nn\\
& &\qquad\left. - \frac{1}{12} \left(e^{-\phi}H_{\mu\nu\lambda}^{(1)} 
H^{(1)\, \mu
\nu\lambda}
+ e^{\phi}\left(H_{\mu\nu\lambda}^{(2)} 
+ \chi H_{\mu\nu\lambda}^{(1)}\right)\left(
H^{(2)\, \mu\nu\lambda} + \chi H^{(1)\,
\mu\nu\lambda}\right)\right)\right]\nn\\
\eqaend
It is to be noted that (12) reduces precisely to the effective action
(1) when the RR fields are set to zero. Since (1) is a special case of 
(12), the five-brane solution obtained from (1) (given in (6), (8) or (9))
can also be generalized for the type IIB theory. We are going to construct 
in the following these generalized solution of five-branes of type IIB
theory. The construction will be facilitated if we write the action (12)
in the manifestly SL(2, R) invariant form as given below [12,19]:
\begineq
{S}_{{\rm IIB}}\,\,=\,\, \int\,d^{10} x \sqrt {- g}
\left[R + \frac{1}{4} {\rm tr}\, \nabla_{\mu} {\cal M}
\nabla^{\mu}{\cal M}^{-1} - \frac{1}{12}{\cal H}_{\mu\nu
\lambda}^T {\cal M} {\cal H}^{\mu\nu\lambda}\right]
\endeq 
where
${\cal M}\,\,\equiv\,\, \left(\begin{array}{cc}
\chi^2 + 
e^{- 2{\phi}}
&  \chi \\
 \chi  &  1\end{array}\right)\,\,e^{{\phi}}$ represents an SL(2, R)
matrix and ${\cal H}_{\mu\nu\lambda} \equiv \left(\begin{array}{c}
H_{\mu\nu\lambda}^{(1)} \\ H_{\mu\nu\lambda}^{(2)}\end{array}\right)$. Also
the superscript `$T$' represents the transpose of a matrix. The action (13)
can be easily seen to be invariant under the following global SL(2, R)
transformations:
\begineq
{\cal M} \rightarrow \Lambda {\cal M} \Lambda^T, \qquad 
{\cal H}_{\mu\nu\lambda} \rightarrow \left(\Lambda^{-1}\right)^T {\cal H}_{
\mu\nu\lambda}, \qquad g_{\mu\nu} \rightarrow g_{\mu\nu}
\endeq
where $\Lambda = \left(\begin{array}{cc} a & b\\ c & d\end{array}\right)$, 
with
$ad - bc = 1$, represents a global SL(2, R) transformation matrix. It is
easy to check that under the transformation (14), the complex scalar
$\lambda = \chi + i e^{-\phi}$ and the two field strengths $H_{\mu\nu\lambda}
^{(1)}$ and $H_{\mu\nu\lambda}^{(2)}$ transform as,
\eqabegin
\lambda &\rightarrow& \frac{a\lambda + b}{c\lambda + d}\nn\\
H_{\mu\nu\lambda}^{(1)} &\rightarrow& d H_{\mu\nu\lambda}^{(1)} - c 
H_{\mu\nu\lambda}^{(2)}\nn\\
H_{\mu\nu\lambda}^{(2)} &\rightarrow& - b H_{\mu\nu\lambda}^{(1)} + a 
H_{\mu\nu\lambda}^{(2)}
\eqaend
We would like to point out here that unlike the electrically charged
string solution, the magnetic charges associated with 
$H_{\mu\nu\lambda}^{(1)}$ 
and $H_{\mu\nu\lambda}^{(2)}$ of the five-brane should transform in the
same way as the field strengths themselves. This follows from the fact
that Noether charge (or the electrical charge) of the string solution
is conserved due to the equation of motion following from (13) 
whereas the topological charge (or the magnetic charge) of the five-brane
is conserved due to Bianchi identity. Therefore the magnetic charges
of the five-branes transform as ${\cal Q} \rightarrow 
(\Lambda^{-1})^T {\cal Q}$        
or
in components,
\eqabegin
Q^{(1)} &\rightarrow& d Q^{(1)} - c Q^{(2)}\nn\\
Q^{(2)} &\rightarrow& -b Q^{(1)} + a Q^{(2)}
\eqaend
Note that the original solution (8) or (9) had one charge $Q$ associated
with $H^{(1)} = Q \epsilon_3$ and this charge was quantized in some
basic units. Now after the transformation (16) $Q$ will no longer remain
quantized. So, in order to recover the charge quantization [20] we
modify the original charge by $\Delta_{(q_1, q_2)}^{1/2} Q$, where
$\Delta_{(q_1, q_2)}$ is an arbitrary constant which will be determined
later. The construction of the SL(2, R) matrix $\Lambda$ can be motivated
such that it properly reproduces the asymptotic value of the complex
scalar $\lambda_0 = \chi_0 + i e^{-\phi_0}$, after the transformation,
where $\phi_0$ and $\chi_0$ are the asymptotic value of the dilaton and 
the RR scalar. The relevant SL(2, R) transformation matrix then takes 
the form [12,17]
\begineq
\Lambda =
\left(\begin{array}{cc} 
e^{-\phi_0} \cos \alpha + \chi_0 \sin \alpha & - e^{-\phi_0} \sin \alpha
+ \chi_0 \cos \alpha\\
\sin \alpha & \cos \alpha\end{array}\right)\,e^{\phi_0/2}
\endeq
Here $\alpha$ is an arbitrary parameter which will be fixed from the
charge quantization condition. Then from (16), we find the charges
associated with $H_{\mu\nu\lambda}^{(1)}$ and $H_{\mu\nu\lambda}^{(2)}$
to be given as,
\eqabegin
Q^{(1)} &=& e^{\phi_0/2} \cos \alpha 
\Delta_{(q_1, q_2)}^{1/2} Q\nn\\
Q^{(2)} &=& \left(e^{-\phi_0/2} \sin \alpha - \chi_0 e^{\phi_0/2}
\cos \alpha\right) \Delta_{(q_1, q_2)}^{1/2} Q
\eqaend
By demanding that the charges be quantized we find,
\eqabegin
\cos \alpha &=& e^{-\phi_0/2} \Delta_{(q_1, q_2)}^{-1/2} q_1\nn\\
\sin \alpha &=& e^{\phi_0/2} \left(q_2 + q_1 \chi_0\right) \Delta_
{(q_1, q_2)}^{-1/2}
\eqaend
where $q_1$ and $q_2$ are integers. Using $\cos^2 \alpha + \sin^2 \alpha
= 1$, (19) determines the value of $\Delta_{(q_1, q_2)}$ as,
\eqabegin
\Delta_{(q_1, q_2)} &=& e^{-\phi_0} q_1^2 + \left(q_2 + q_1 \chi_0\right)
^2 e^{\phi_0}\nn\\
&=& \left(q_1,\,\, q_2\right) {\cal M}_0 \left(\begin{array}{c}
q_1 \\ q_2\end{array}\right)
\eqaend
where ${\cal M}_0 = \left(\begin{array}{cc} \chi_0^2 + e^{-2\phi_0}
& \chi_0\\ \chi_0 & 1\end{array}\right)\,e^{\phi_0}$. It is clear from (20)
that $\Delta_{(q_1, q_2)}$ is SL(2, Z) covariant. Therefore, the charge of
a $(q_1, q_2)$ five-brane is given by an SL(2, Z) covariant expression
\eqabegin
Q_{(q_1, q_2)} &=& \Delta_{(q_1, q_2)}^{1/2} Q\nn\\
&=& \sqrt{e^{-\phi_0} q_1^2 + \left(q_2 + q_1 \chi_0\right)^2 
e^{\phi_0}}~Q
\eqaend
Note from (14) that the canonical metric does not change under the
SL(2, R) transformation. However, since the charge $Q$ is now replaced
by $Q_{(q_1, q_2)}$ the metric given in (9) takes the following
form:
\begineq
ds^2 = \left(1 + \frac{Q_{(q_1, q_2)}}{2 \rho^2}\right)^{-1/4}
\left[ -(dt)^2 + \delta_{ij}dx^i dx^j\right] + \left(1 + \frac{
Q_{(q_1, q_2)}}{2\rho^2}\right)^{3/4}\left(d\rho^2 + \rho^2
d\Omega_3^2\right)
\endeq
The complex scalar field $\lambda$ changes as
\eqabegin
\lambda &=& \frac{a\left(i e^{-\phi}\right) + b}
{c \left(i e^{-\phi}\right) + d}\nn\\
&=& \frac{\chi_0 \Delta_{(q_1, q_2)} A_{(q_1, q_2)} + q_1 q_2 e^{-\phi_0}
\left(A_{(q_1, q_2)} - 1\right) + i \Delta_{(q_1, q_2)} A_{(q_1, q_2)}^
{1/2} e^{-\phi_0}}{q_1^2 e^{-\phi_0} + A_{(q_1, q_2)} e^{\phi_0} \left(
\chi_0 q_1 + q_2\right)^2}
\eqaend
where $A_{(q_1, q_2)} = \left(1 + \frac{Q_{(q_1, q_2)}}{2\rho^2}\right)^
{-1}$. Note that asymptotically as $\rho \rightarrow \infty$, $A_{(q_1,
q_2)} \rightarrow 1$ and therefore, $\lambda \rightarrow \lambda_0$ 
as expected. The real and imaginary part of (23) give the transformed value
of the RR scalar and the dilaton of the theory. Finally, the 
transformed value of the field strengths $H^{(1)}$ and $H^{(2)}$ can
be obtained from (15) as,
\eqabegin
H^{(1)} &=& q_1 Q \epsilon_3\nn\\
H^{(2)} &=& q_2
Q\epsilon_3
\eqaend
which can be written compactly as follows,
\begineq
{\cal H} = \left(\begin{array}{c} q_1\\ q_2\end{array}\right)
Q \epsilon_3
\endeq
We can also calculate
the tension of a $(q_1, q_2)$ five-brane by calculating the ADM
mass per unit five-volume [21]. We note that in general the ADM
mass is given by $M = (3 r_+^2 - r_-^2)$ and therefore, for the
non-extremal case the mass and the charge are independent parameters.
But in the extremal case they are related since in that case,
$M = 2 r_-^2 = Q$. We have seen in (21) that the charge of a $(q_1,
q_2)$ five-brane has been modified by $\Delta_{(q_1, q_2)}^{1/2} Q$
and so in order to equate the mass with the charge, mass per unit
five-volume i.e. the tension must also satisfy the similar relation:
\eqabegin
T_{(q_1, q_2)} &=& \Delta_{(q_1, q_2)}^{1/2} T\nn\\
&=&\sqrt{e^{-\phi_0} q_1^2 + \left(q_2 + q_1 \chi_0\right)^2 e^{\phi_0}} 
~T\nn\\
&=&\sqrt{g_s^{-1} q_1^2 + \left(q_2 + q_1 \chi_0\right)^2 g_s}~ T
\eqaend
where $g_s = e^{\phi_0}$ in the last expression of (26) denotes the string
coupling constant. 
Thus when
$\chi = 0$, the solitonic five-brane or (1, 0) brane tension is
proportional to $1/\sqrt{g_s}$ whereas D5-brane or (0, 1)
brane tension is proportional to $\sqrt{g_s}$ in the canonical
metric. In the string metric, on the other hand, the tension of
a general $(q_1, q_2)$ five-brane is given by,
\begineq
T_{(q_1, q_2)} = g_s^{-3/2} \sqrt{g_s^{-1} q_1^2 + g_s q_2^2}~ T
\endeq
Here $T_{(q_1, q_2)}$ gets scaled by $g_s^{-3/2}$ because, it has the
dimensionality of (length)$^{-6}$.
Thus in the string metric, the tension of a solitonic five-brane
is proportional to $1/g_s^2$ and the tension of a D5-brane is
proportional to $1/g_s$ as expected. This tension formula of a 
$(q_1, q_2)$ five-brane has been envisaged by Witten in ref.[15].

Thus starting from the solitonic five-brane solution, we have 
obtained an infinite family of five-brane solutions permuted
by SL(2, Z) group in type IIB theory given by the metric and 
other field configurations in (22) -- (25).

The stability of these $(q_1, q_2)$ five-brane solutions can
be understood along the same line as in the case of string
solutions [17,21]. Since the tension of a $(q_1, q_2)$ five-brane
is given in (26), it can be easily checked that when $\chi = 0$,
the five-brane tension satisfy the following triangle inequality
\begineq
T_{(p_1, p_2)} + T_{(q_1, q_2)} \geq T_{(p_1+q_1, p_2+q_2)}
\endeq
Such relation is quite typical of a BPS state. The equality
holds when $p_1 q_2 = p_2 q_1$ or in other words when $p_1
= n q_1$ and $p_2 = n q_2$, where $n$ is any integer. Thus
when $q_1$ and $q_2$ are relatively prime, the inequality
prevents the five-brane state to decay into five-branes of
lower masses. Since the charge of a $(q_1, q_2)$ five-brane 
also satisfies similar relation (21) it can be readily checked
again that when $q_1,\,q_2$ are coprime the charge conservation
can not be satisfied if the five-brane decay into multiple
five-branes. Thus $(q_1, q_2)$ five-brane configuration with
$q_1$, $q_2$ relatively prime, describes a bound state
configuration of $q_1$ solitonic five-branes with $q_2$ D5-branes.

Note here, as discussed by Witten, that unlike the string solution,
the D5-branes themselves do not form bound states as the six
dimensional super Yang-Mills theory does not contain vacua with
a mass gap. On the other hand, D5-branes when combined with
solitonic five-branes do form bound states and has been discussed
qualitatively by Witten in ref.[15].

The SL(2, Z) multiplet of black fivebranes can be obtained similarly.
Eqs. (14)--(21) and Eqs. (23)--(25) remain the same but (22) should be replaced 
by (6) with $r_+$ and $r_-$ now given by 
$2 (r_+ r_-) = \Delta_{(q_1, q_2)}^{1/2} Q$. The same procedure applies to the
 construction of the SL(2, Z) multiplet of black strings.  These multiplets
 may be useful in studying the physics of black strings and fivebranes.

Finally, we would like to mention that similar infinite family
of magnetic dual solutions of strings also exist in $ D > 4$. The
low energy effective action common to any string theory in $D$
dimensions has the form:
\begineq
S_D = \int\,d^D x \sqrt {-g} 
\left[R - \frac{4}{D-2} \nabla_\mu \phi \nabla^\mu \phi - \frac{1}{12}
e^{-\frac{8}{D-2}\phi} H_{\mu\nu
\lambda}^{(1)} H^{(1)\,\mu\nu\lambda}\right]
\endeq
The equations of motion following from (29) are given as:
\eqabegin
& & \nabla_\mu\left(e^{-\frac{8}{D-2}\phi} H^{(1)\,\mu\nu\lambda}
\right)\,\,\,
=\,\,\,0\\
& &\nabla^2 \phi + \frac{1}{12} e^{-\frac{8}{D-2}\phi} 
\left(H^{(1)}\right)^2\,\,\,
=\,\,\,0\\
& & R_{\mu\nu} - \frac{4}{D-2} \nabla_\mu \phi \nabla_\nu \phi -
\frac{1}{4} e^{-\frac{8}{D-2}\phi} H_{\mu\rho\sigma}^{(1)} 
H_\nu^{(1)\,\,
\rho\sigma} 
 + \frac{1}{6(D-2)} e^{-\frac{8}{D-2}\phi} 
\left(H^{(1)}\right)^2 
g_{\mu\nu}\,\,\,=\,\,\,0\nn\\
\eqaend
By using the same ansatz on the metric as before one can obtain the
magnetic dual solution of the string in $D > 4$ dimensions of the
form:
\begineq
ds_D^2 = A^{-\frac{2}{D-2}}\left[-(dt)^2 + \delta_{ij} dx^i dx^j\right]
+ A^{\frac{D-4}{D-2}}\left(d\rho^2 + \rho^2 d\Omega_3^2\right)
\endeq
where $A$ is a function of radial coordinate $\rho$ only whose 
explicit form is given in [9] as:
\begineq
e^{2\phi} = \left( A (\rho) \right)^{{\sqrt {D - 2 \over 8}}} = 
\left (1 + \frac{r_-^2}{  \rho^2}\right)^{{\sqrt{D - 2 \over 8}}} 
\endeq
where $r_-$ is the charge of the dual object. Also, the field strength
$H^{(1)} = Q \epsilon_3$. The magnetic dual object is a 4-brane in $D = 9$, 
a 3-brane in $D = 8$, a 2-brane in $D = 7$, a string in $D = 6$ and a 0-brane 
(a particle) in $D = 5$.

Since it is known that the toroidally compactified type IIB string
theory also possesses the SL(2, R) invariance [23,24] (This symmetry can also 
be obtained in $D \leq 9$ from toroidally compactified M-theory, 
see, e.g., [25].) with the
same transformation properties of ${\cal H}_{\mu\nu\lambda}$ and
the complex scalar $\lambda$ and since the reduced action can be
shown to be given by (29) when RR fields are set to zero, we 
can straightforwardly use the SL(2, Z) rotation to find the SL(2, Z)
family of the dual solutions of strings starting from (33) and (34). The
solutions in this case are very similar as in the ten dimensional
case. (For the detailed construction of string solution in $D<10$ see
[17].) The corresponding black SL(2, Z) multiplet for each of the dual objects 
can be constructed in a similar fashion described above.

To conclude, we have constructed in this paper an SL(2, Z) family
of super five-brane solutions in type IIB string theory starting from
the known non-singular solitonic five-brane solution. These 
solutions are characterized by a pair of integers corresponding to the 
magnetic charges associated with the two three-form field strengths
in the NSNS and RR sector of the theory. We have shown that both
the charge and the tension of a general $(q_1, q_2)$ five-brane are
given by SL(2, Z) covariant expressions. This provides more evidence 
in support 
of the conjecture that SL(2, Z) is an exact symmetry group of the
quantum type IIB string theory. When the integers $q_1$ and $q_2$
are relatively prime, we have shown that the five-brane is stable
since it is prevented from decaying into multiple five-branes by
a triangle inequality relation of both the tension as well as the
charge associated with the five-brane. We have obtained the tension 
formula for a general $(q_1, q_2)$ five-brane as envisaged by 
Witten in ref.[15]. We have discussed that a similar family of
the magnetic-dual solutions of the string also exists in each of $D > 4$ 
dimensions in type II string theory. We also discussed how to obtain the 
corresponding SL(2, Z) multiplets for black strings and its dual objects for
$ 10 \geq D > 4$.    

\vspace{1cm}

\begin{large}
\noindent{\bf References:}
\end{large}

\vspace{.5cm}

\begin{enumerate}

\item G. W. Gibbons, \np 207 (1982) 337; G. W. Gibbons and K. Maeda,
\np 298 (1988) 741.
\item D. Garfinkle, G. Horowitz and A. Strominger, Phys. Rev. D43 (1991)
3140, Erratum: Phys. Rev. D45 (1992) 3888.
\item G. Horowitz and A. Strominger, \np 360 (1991) 197.
\item A. Dabholkar and J. A. Harvey, Phys. Rev. Lett. 63 (1989) 478;
A. Dabholkar, G. Gibbons, J. A. Harvey and F. Ruiz Ruiz, \np 340
(1990) 33.
\item A. Strominger, \np 343 (1990) 167.
\item S. J. Rey, Phys. Rev. D43 (1991) 526.
\item C. G. Callan, J. A. Harvey and A. Strominger, \np 359 (1991) 611.
\item M. J. Duff and J. X. Lu, \np 354 (1991) 141.
\item M. J. Duff, R. R. Khuri and J. X. Lu, Phys. Rep. 259 (1995) 213.
\item B. Julia, in {\it Supergravity and Superspace}, Eds. S. W.
Hawking and M. Rocek, Cambridge University Press, 1981.
\item C. M. Hull and P. K. Townsend, \np 438 (1995) 109.
\item J. H. Schwarz, \pl 360 (1995) 13 (see revised version,
hep-th/9508143).
\item E. Bergshoeff, H. J. Boonstra and T. Ortin, Phys. Rev. D53 (1996)
7206. 
\item R. Poppe and S. Schwager, \pl 393 (1997) 51.
\item E. Witten, \np 460 (1996) 335.
\item J. Polchinski, Phys. Rev. Lett. 75 (1995) 4724; J. Polchinski,
S. Chaudhuri and C. V. Johnson, {\it Notes on D-Branes}, hep-th/9602052;
J. Polchinski, {\it TASI Lectures on D-Branes}, hep-th/9611050.
\item S. Roy, {\it SL(2, Z) Multiplets of Type II Superstrings in
$D < 10$}, hep-th/9706165 (to appear in Phys. Lett. B).
\item M. J. Duff and J. X. Lu, \pl 273 (1991) 409. 
\item C. Hull, \pl 357 (1995) 545.
\item R. I. Nepomechie, Phys. Rev. D31 (1985) 1921; C. Teitelboim,
\pl 167 (1986) 69.
\item J. X. Lu, \pl 313 (1993) 29. 
\item J. H. Schwarz, {\it Lectures on Superstring and M-Theory
Dualities}, hep-th/9607201.
\item J. Maharana, \pl 402 (1997) 64.
\item S. Roy, {\it On S-Duality of Toroidally Compactified Type
IIB String Effective Action}, hep-th/9705016.
\item  E. Cremmer, B. Julia, H. Lu and C. Pope, {\it Dualisation of Dualities. 
I.}, hep-th/9710119.

\end{enumerate}

\vfil\eject
\end{document}